\documentclass[prb,floatfix,aps,twocolumn,showpacs]{revtex4-1}
\usepackage[dvips]{graphics}
\usepackage{graphicx}
\usepackage[utf8]{inputenc}
\usepackage{amssymb}
\usepackage{epstopdf}
\usepackage{bm}
\usepackage{dcolumn}
\usepackage{epsfig}
\usepackage{latexsym}
\usepackage{amsmath}
\usepackage{color}
\usepackage{array}
\usepackage{enumerate,dsfont}
\usepackage{xr}
\def\XXint#1#2#3{{\setbox0=\hbox{$#1{#2#3}{\int}$ }
\vcenter{\hbox{$#2#3$ }}\kern-.5\wd0}}

\renewcommand{\vec}[1]{\bm{#1}}
\newcommand{\re}[1]{\mathrm{Re}({#1})}
\newcommand{\ket}[1]{\left |\mbox{$#1$}\right\rangle}
\newcommand{\bra}[1]{\left\langle\mbox{$#1$}\right |}

\newcommand{\ch}{\mathrm{ch}}
\newcommand{\angstrom}{\textup{\AA}}
 
\begin{document}
\title{Long-Distance Entanglement of Spin-Qubits via  Ferromagnet}
\author{Luka Trifunovic, Fabio L. Pedrocchi, and Daniel Loss}
\affiliation{Department of Physics, University of Basel, Klingelbergstrasse 82, CH-4056 Basel, Switzerland}
\begin{abstract}
We propose a mechanism of coherent coupling between distant spin qubits
interacting dipolarly with a ferromagnet. We derive an effective two-spin
interaction Hamiltonian and estimate the coupling strength. We discuss the
mechanisms of decoherence induced solely by the coupling to the ferromagnet and
show that there is a regime where it is negligible.  Finally, we present a
sequence for the implementation of the entangling CNOT gate and estimate the
corresponding operation time to be a few tens of nanoseconds.  A particularly
promising application of our proposal is to atomistic spin-qubits such as
silicon-based qubits and NV-centers in diamond to which existing coupling
schemes do not apply.
\end{abstract}
\maketitle

\section{Introduction}
Quantum coherence and entanglement lie at the heart of quantum information
processing. One of the basic requirements for implementing quantum computing is
to generate, control, and, measure entanglement in a given quantum system. This
is a rather challenging task, as it requires to overcome several obstacles, the
most important one being  decoherence processes. These negative effects have
their origin in the unavoidable coupling of the quantum systems to the
environment they are residing in. 

A guiding principle in the search for a good system to encode qubits is the
smaller the system the more coherence, or, more precisely, the fewer degrees of
freedom the weaker the coupling to the environment. Simultaneously, one needs to
be able to coherently manipulate the individual quantum objects, which is more
efficient for larger systems. This immediately forces us to compromise between
manipulation and decoherence requirements.

Following this principle, among the most promising candidates for encoding a
qubit we find \emph{atomistic} two-level systems, such as NV-centers and
silicon-based spin
qubits.~\cite{jelezko_observation_2004,jelezko_observation_2004-1,Childress13102006,Dutt01062007,
  Hanson18042008,Neumann06062008,Jiang09102009,Fuchs11122009,Neumann30072010,
Buckley26112010,togan2010quantumentanglement,robledo2011highfidelityprojective}
The latter are composed of nuclear (electron) spins of phosphorus atoms in a
silicon nanostructure. They have very long $T_2$ times of
$60\,ms$~\cite{pla_high_2013} for nuclei and of $200\mu s$ for
electrons.~\cite{pla_single-atom_2012} Recently, high fidelity single qubit
gates and readout have been demonstrated
experimentally.~\cite{pla_single-atom_2012} Nitrogen-vacancy
centers~\cite{dobrovitski_quantum_2013} in diamond have also been demonstrated
experimentally to be very stable with long decoherence times of
$T_{2}^{*}\approx 20\,\mu s$ and $T_{2}\approx
1.8\,ms$.~\cite{balasubramanian2009ultralongspin} Both types of spin qubits
have the additional advantage that noise due to surrounding nuclear spins can
be avoided by isotopically purifying the material. 

Unfortunately, it is hardly possible to make these spin qubits interact with
each other in a controlled and scalable fashion. They are very localized and
their position in the host material is given and cannot be adjusted easily.
Therefore, if during their production two qubits turn out to lie close to each
other they will always be coupled, while if they are well-isolated from each
other they will never interact. It is thus of high interest to propose a scheme
to couple such atomistic qubits in a way that allows a high degree of control. 

\begin{figure}[ht]
  \centering\includegraphics[width=\columnwidth]{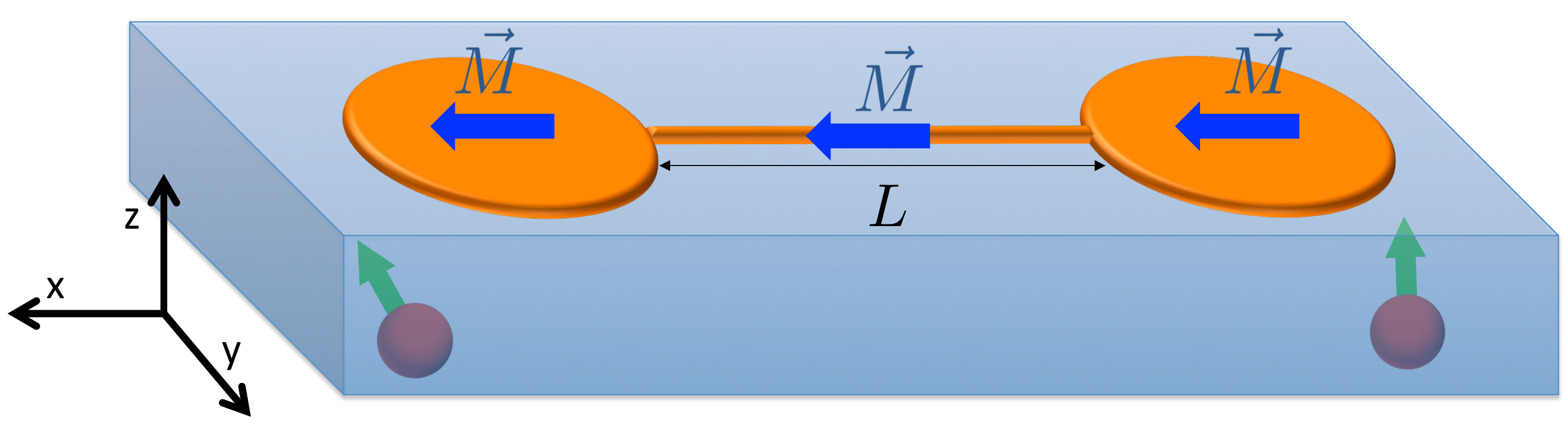}
  \caption{The schematics of the ferromagnetic coupler setup. The orange
    dog-bone shape denotes the ferromagnet that is coupled via magnetic dipole interaction to 
    spins of nearby quantum dots (red sphere with green arrow). The ferromagnet is
    assumed to be a monodomain and its magnetization is denoted by blue arrows
    ($\vec M$) that can take arbitrary orientation. $L$ is the length of the
    quasi-1D ferromagnetic channel that is approximately equal to the distance
    between the qubits. The shape of the ferromagnetic coupler is chosen such
    that it enables strong coupling to the spin-qubits while maintaining the
    spatially slowly decaying 1D susceptibility between the two discs.}
  \label{fig:setup_fig}
\end{figure}

We fill this gap in the present work by proposing a setup to couple two spin
qubits separated by a relatively large distance on the order of micrometers, see
Fig.~\ref{fig:setup_fig}. The coupling is mediated via a ferromagnet with gapped
excitations to which the spin qubits are coupled by magnetic dipole-dipole
interaction. Since the ferromagnet is gapped only virtual magnons are excited
but in order to obtain the sizable coupling one needs to tune the splitting of
the qubit close to resonance with the gap of the ferromagnet. The \textit{on}
and \textit{off} switching of the qubit-qubit interaction is therefore achieved
by tuning qubits off resonance (see below). The resulting system is thus
realizable with present state-of-the-art technologies.  We point out that our
analysis is not restricted to a precise type of spin qubit but is in principle
applicable to any system that dipolarly interact with the spins of a
ferromagnet. In particular, our proposal is also applicable to an electron spin
localized in a semiconductor quantum dot, gate-defined or
self-assembled.~\cite{loss_quantum_1998,kloeffel_prospects_2013} While 
other schemes exist to couple such qubits over large
distances,~\cite{trifunovic_long-distance_2012,shulman_demonstration_2012,childress_mesoscopic_2004,burkard_2006,
trif_spindyn_2008} none of them is applicable to atomistic qubits. The main
novelty of our proposal is thus the possibility to \emph{also} couple atomistic
qubits that are of high technological relevance.

Before we proceed with the quantitative analysis, let us first give an
intuitive picture of the qubit-qubit coupling. The coupling between two distant
qubits is mediated via a \textit{coupler} system. The relevant quantity of this
coupler is its spin-spin susceptibility---in order to have a long-range
coupling, a slowly spatially decaying susceptibility is required. The
dimensionality of the coupler plays an important role since, in general, it
strongly influences the spatial decay of the susceptibility, which can be
anticipated from purely geometric considerations. Furthermore, since the
coupler interacts with the qubits via magnetic dipolar forces, we require that
a large part of the coupler lies close to the qubits. To this end we
immediately see that a dog-bone shape depicted in Fig.~\ref{fig:setup_fig}
satisfies these two requirements---strong dipolar coupling to the qubits and
slow spatial, practically 1D, susceptibility decay between the qubits.

\section{Model}
The system we consider consists of two spin-$\frac{1}{2}$ qubits coupled
dipolarly to the ferromagnet
\begin{equation}
  H=H_\sigma+H_{\rm F}+H_{\rm I},
  \label{eq:Hgen}
\end{equation}
where $H_F$ is for the moment unspecified Hamiltonian of the dog-bone shaped ferromagnet that is
assumed to be polarized along the $x$-axis. We first assume that the qubits are also
polarized along the $x$-axis,
$H_\sigma=\sum_{i=1,2}\frac{\Delta_i}{2}\sigma_i^x$, while the ferromagnet disc
axes are along $z$, see Fig.~\ref{fig:setup_fig}. The magnetic dipole coupling between
the ferromagnet and the spin-qubits can be written as
\begin{eqnarray}
  H_{\rm I}&=&\frac{\mu_0\mu_b\mu}{4\pi a^3}\sum_{i=1,2}\int d\bm r S_{\bm r}^x\left[ \left(
  \frac{3iA_{i,\bm r}^{\prime}}{2}+\frac{3C_{i,\bm r}^{\prime\prime}}{4}
  \right)\sigma_i^++\text{h.c.}\right.\nonumber\\
  &&\hspace{4cm}+\left.\frac{1}{2}\left( B_{i,\bm r}-3C_{i,\bm r}^\prime
  \right)\sigma_i^z\right]\nonumber\\
  &&+S_{\bm r}^+\left[\left(\frac{3}{8}C_{i,\bm r}^{\prime}-\frac{3i}{2}A_{i,\bm r}^{\prime\prime}
  +\frac{3}{8}B_{i,\bm r} \right)\sigma_i^+\right.\nonumber\\
 && -\frac{1}{8}\left.\left(B_{i,\bm r}-3C_{i,\bm r}^{\prime}
  \right)\sigma_i^-+\left( \frac{3C_{i,\bm r}^{\prime\prime}}{4}+\frac{3iA_{i,\bm r}^{\prime}}{2}
  \right)\sigma_i^z\right]\nonumber\\
 &&+\text{h.c},
  \label{eq:Dipolar}
\end{eqnarray}
where $A_{\bm r}$, $B_{\bm r}$, $C_{\bm r}$ are given by
\begin{align}
  A_{\bf r}&=\frac{1}{a^{3}}\frac{r^{z}r^{+}}{r^5}\,,\label{eq:A}\\
  C_{\bf r}&=\frac{1}{a^{3}}\frac{(r^+)^2}{r^5}\,,\label{eq:C}\\
  B_{\bf
  r}&=\frac{1}{a^{3}}\frac{1}{r^3}\left(2-\frac{3r^+r^-}{r^2}\right)\,\label{eq:B},
\end{align}
with $S_{\bm r}^{\pm}=S_{\bm r}^{y}\pm iS_{\bm r}^{z}$ and lattice constant $a$. Here we denote the real part of a complex number with prime and the imaginary part with double prime. The operator $\bm
S_{\bf r}$ describes the spin of the ferromagnet at the position $\bf r$.

Next, we release the assumptions about the mutual orientation of the disc axes, the
axes of polarization of the ferromagnet, and the direction of the qubits splitting and assume that these
can take arbitrary directions. Now the interaction Hamiltonian reads
\begin{align}
  H_{\rm I}=&\frac{\mu_0\mu_b\mu}{4\pi a^3}\sum_{i=1,2}\int d\bm r S_{\bm r}^{\tilde z}\left[ 
  a_{i,\bm r}\sigma_i^z+b_{i,\bm r}\sigma_i^++\text{h.c.}\right]+\nonumber\\
  &S_{\bm r}^{\tilde +}\left[ c_{i,\bm r}\sigma_i^z+d_{i,\bm r}\sigma_i^++ e_{i,\bm r}\sigma_i^-
  \right]+\text{h.c},
  \label{eq:DipolarGen}
\end{align}
where $\bm S_{\bm r}$ and $\bm \sigma_{\bm r}$ have, in general, different quantization axes. The
expressions of the coefficients in Eq.~(\ref{eq:DipolarGen}) are now more
complicated, nevertheless it is important to note that the integrals of these
coefficients are \emph{experimentally accessible}. The qubits can be used to measure the
stray field of the ferromagnet which is given by
$B_s=(b_i^\prime,b_i^{\prime\prime},a_i)$, where
$\{a_i,\dots,e_i\}=\frac{\mu_0\mu}{4\pi a^3}\int d\bm r \{a_i,\dots,e_i\}_{\bm
r}$. In order to measure the remaining coefficients, one needs to apply the
magnetic field externally in order to polarize sequentially the ferromagnet
along the two perpendicular directions to the ferromagnet easy axis. The coefficients are obtained then by measuring again the stray fields (with the aid of the
qubits) which now are given by
$(d^\prime+e^\prime,d^{\prime\prime}-e^{\prime\prime},c^\prime)$ and
$(d^{\prime\prime}+e^{\prime\prime},d^\prime-e^\prime,c^{\prime\prime})$.
Furthermore, all the results that we are going to obtain for the qubit-qubit
coupling as well as the estimates of the decoherence will depend only on the
integrals of the coefficients, i.e., on $\{a_{i},\dots,e_{i}\}$ rather than $\{a_{i},\dots,e_{i}\}_{\bm r}$.

\subsection{Coherent coupling}
We proceed to derive the effective qubit-qubit coupling by performing a
Schrieffer-Wolff (SW) transformation.~\cite{bravyi_schrieffer_2011} We assume
that the excitations in the ferromagnet are gapped due to some magnetic
anisotropy (e.g.~shape-anisotropy), with the gap being denoted by $\Delta_F$.
This is important because when the qubit splitting $\Delta$ is smaller than
$\Delta_F$, flipping the qubit spin cannot excite magnons in the ferromagnet,
thus there are only virtual magnons excited via coupling to the
qubits---otherwise such a coupling would lead to strong decoherence in the
qubits. Due to the presence of the gap in the ferromagnet, its transversal
susceptibility $\chi_\perp(\omega,\bm r)$ decays exponentially for
$\omega<\Delta_F$ with the characteristic length
$l_F\propto1/\sqrt{\Delta_F-\omega}$, thus we take into account only terms with
$\omega\sim\Delta_F$, see Appendix. Straightforward application of lowest order SW transformation
accompanied by tracing out the degrees of freedom of the ferromagnet yields the effective
qubit-qubit coupling Hamiltonian 
\begin{align}
  H_{\rm eff}&=H_\sigma+\chi_\perp^{\rm
  1D}(\Delta_1,L)e_1\sigma_1^-(c_2\sigma_2^z+d_2\sigma_2^++e_2\sigma_2^-)^\dagger+\nonumber\\
  &1\leftrightarrow2+\text{h.c.},
  \label{eq:HeffGen}
\end{align}
where $\chi_\perp^{\rm 1D}$ is the transverse susceptibility (i.e. transverse to the $\widetilde z$ direction) of a quasi-1D ferromagnet,
since we assumed a dog-bone shaped ferromagnet.  We have neglected the
longitudinal susceptibility $\chi_\parallel$ since it is smaller by factor of
$1/S$ compared to the transverse one and it is suppressed by temperature. It is readily seen from the
above expression that in order to obtain a sizable coupling between the qubits we
have to tune at least one of the qubits close to resonance,
$\Delta_i\sim\Delta_F$. This can be achieved by conveniently positioning the qubit such
that the Zeeman splitting produced by the stray field of the ferromagnet is
close to the excitation gap of the ferromagnet. The fine tuning can be then
achieved by applying locally a small external magnetic field from a coil.
The on resonance requirement offers an elegant way to switch on/off the coupling between the qubits. The idea is to tune the qubit splitting close to resonance to switch on the mediated interaction and to tune it off resonance to switch off the mediated interaction. \footnote{Another possibility is to keep one of the qubits off resonance and then tuning the other one on and off.} 

For the sake of completeness, in the Appendix we present a detailed discussion of the effective coupling 
mediated by the dog-bone when the qubits are exchange coupled to the ferromagnet which requires
a tunnel coupling between spin qubit and ferromagnet.

\subsection{Implementation of two-qubit gates}
Two qubits interacting via the ferromagnet evolve according to the Hamiltonian
$H_\mathrm{eff}$, see Eq.~(\ref{eq:HeffGen}). The Hamiltonian is therefore the sum
of Zeeman terms and qubit-qubit interaction. These terms, by and large, do not
commute, making it difficult to use the evolution to implement standard
entangling gates. Nevertheless, if we assume that $\Delta_1=\Delta_2$,
$H_\sigma$ acts only in the subspace spanned by
$\{\ket{\uparrow,\uparrow},\ket{\downarrow,\downarrow}\}$ and the Zeeman
splitting of the qubits is much larger than the effective qubit-qubit coupling,
we can neglect the effect of $H_\mathrm{eff}$ in this part of the subspace and
approximate it by its projection in the space spanned by the vectors
$\{\ket{\uparrow,\downarrow},\ket{\downarrow,\uparrow}\}$
\begin{equation}
  H'_\mathrm{eff}=H_\sigma+\alpha(\sigma_1^x\sigma_2^x-\sigma_1^y\sigma_2^y)+\beta(\sigma_1^x\sigma_2^y+\sigma_1^y\sigma_2^x),
  \label{eq:Heffprime}
\end{equation}
where $\alpha=-8\re{e_1e_2^*}$ and $\beta=-4\re{d_1e_2^*+d_2e_1^*}$.
Within this approximation, the coupling in $H'_\mathrm{eff}$ and the Zeeman terms
now commute. From here we readily see that the stray field components,
$a_i,b_i$, as well as the coefficient $c_i$ do not determine the operation time
of the two qubit gates---the operation time depends only on $d_i$ and $e_i$. To
proceed we perform a rotation on the second qubit around the $z$-axis by an
angle $\tan\theta=\beta/\alpha$ and arrive at the Hamiltonian
\begin{equation}
  H'_\mathrm{eff}=H_\sigma+\sqrt{\alpha^2+\beta^2}(\sigma_1^x\tilde\sigma_2^x-\sigma_1^y\tilde\sigma_2^y).
  \label{eq:Heffprimerot}
\end{equation}
We consider the implementation of the iSWAP gate
$U_\mathrm{iSWAP}=e^{i(\sigma_1^x\tilde\sigma_2^x+\sigma_1^y\tilde\sigma_2^y)\pi/4}$,
which can be used to implement the CNOT gate.~\cite{nielsen_quantum_2004} The
Hamiltonian $H'$ can be transformed to the desired form by changing the sign of
$\sigma^x_1 \tilde\sigma^x_2$ term. This is achieved with the following
sequence~\cite{imamoglu_quantum_1999}
\begin{equation}
  U_\mathrm{iSWAP}=\sigma_1^ye^{iH_\sigma
  t}e^{-iH'_\mathrm{eff}t}\sigma_1^y,
\end{equation}
where $t = \pi/(4 \sqrt{\alpha^2+\beta^2})$. When iSWAP is available, the CNOT
gate can be constructed in the standard way~\cite{tanamoto_efficient_2008}
\begin{equation}
U_{\rm CNOT}=e^{-i\frac{\pi}{4}\sigma_{1}^z}
e^{i\frac{\pi}{4}\sigma_{2}^x}
e^{i\frac{\pi}{4}\sigma_{2}^z} 
U_{\rm iSWAP} e^{-i\frac{\pi}{4}\sigma_{1}^x} U_{\rm iSWAP} 
e^{i\frac{\pi}{4}\sigma_{2}^z}.
\end{equation}

Since $H'_\mathrm{eff}$ is an approximation of $H_\mathrm{eff}$, the above
sequence will yield approximate CNOT, $U'_{\rm CNOT}$, when used with the full
the Hamiltonian. The success of the sequences therefore depends on the fidelity of
the gates, $F(U'_{\rm CNOT})$. Ideally this would be defined using a
minimization over all possible states of two qubits. However, to characterize
the fidelity of an imperfect CNOT it is sufficient to consider the following
four logical states of two qubits:~\cite{trifunovic_long-distance_2012}
$\ket{+,\uparrow},\ket{+,\downarrow},\ket{-,\uparrow},$ and
$\ket{-,\downarrow}$. These are product states which, when acted upon by a
perfect CNOT, become the four maximally entangled Bell states
$\ket{\Phi^+},\ket{\Psi^+},\ket{\Phi^-},$ and $\ket{\Psi^-}$, respectively. As
such, the fidelity of an imperfect CNOT may be defined,
\begin{equation}
F(U'_{\rm CNOT}) = \min_{i \in \{+,-\}, j \in \{0,1\}} |\bra{i,j} U_{\rm
CNOT}^{\dagger} U'_{\rm CNOT} \ket{i,j}|^2.
\end{equation}
The choice of basis used here ensures that $F(U'_{\rm CNOT})$ gives a good
characterization of the properties of $U'_{\rm CNOT}$ in comparison to a perfect
CNOT, especially for the required task of generating entanglement. For realistic
parameters, with the Zeeman terms two order of magnitude stronger than the
qubit-qubit coupling, the above sequence yields fidelity for the CNOT gate of
$99.976\%$.

To compare these values to the thresholds found in schemes for quantum
computation, we must first note that imperfect CNOTs in these cases are usually
modeled by the perfect implementation of the gate followed by depolarizing
noise at a certain probability. It is known that such noisy CNOTs can be used
for quantum computation in the surface code if the depolarizing probability is
less than $1.1\%$.~\cite{fowler:10} This corresponds to a fidelity, according
to the definition above, of $99.17\%$. The fidelities that may be achieved in
the schemes proposed here are well above this value and hence, though they do
not correspond to the same noise model, we can expect these gates to be equally
suitable for fault-tolerant quantum computation.
\section{Decoherence}
In this section we study the dynamics of a single qubit coupled to the
ferromagnet. In particular we want to answer the question whether the effective
coupling derived in the previous section is \textit{coherent}, i.e., whether the
decoherence time solely due to the dipolar coupling to the ferromagnet is larger
than the qubit operation time.

A ferromagnet has two types of fluctuations---longitudinal and transverse ones.
The longitudinal noise stems from fluctuations of the longitudinal $S^{\tilde
z}$ component (we recall that the ferromagnet is polarized along $\tilde z$),
while the transverse one is related to fluctuations of $S^{\tilde \pm}$. In what
follows we study these two noise sources separately. The general noise model
that describes both types of noise is then given by
\begin{align}
  H=H_{\rm F}+\frac{\Delta}{2}\sigma^z+\sigma^z\otimes X+\sigma^+\otimes
  Y+\text{h.c.},
  \label{eq:NoiseModel}
\end{align}
where the ferromagnet operators $X$ ($Y$) with zero expectation value couple
longitudinally (transversally) to the qubit. The noise model given in
Eq.~(\ref{eq:NoiseModel}) leads to the following relaxation and decoherence
times within Born-Markov approximation~\cite{divincenzo_rigorous_2005}
\begin{align}
  T_1^{-1}&=S_{Y}(\omega=\Delta),\\
  T_2^{-1}&=\frac{1}{2}T_1^{-1}+S_{X}(\omega=0),
  \label{eq:T1T2}
\end{align}
where we defined the fluctuation power spectrum of an operator $A$ in the
following way, $S_A(\omega)=\int dt e^{-i\omega t}\{A^\dagger(t),A(0)\}$. In
order to obtain estimates for the decoherence times we need a specific model
for the ferromagnet Hamiltonian, herein taken to be a gapped Heisenberg model
$H_{F}=-J\sum_{\langle {\bf r},{\bf r'}\rangle}{\bm S}_{{\bf r}}\cdot{\bm
S}_{{\bf r'}}+\Delta_F\sum_{\bf r}S_{\bf r}^z$, $J$ being the exchange coupling
and $\Delta_F$  the excitation gap induced by some magnetic
anisotropy.

\subsection{Longitudinal noise}
The power spectrum of longitudinal fluctuations is given by the following
expression (see Appendix)
\begin{align}
  S_\parallel^{\rm
  3D}(\omega)&=\frac{\alpha\sqrt{\beta\omega}}{2\beta^2D^3}e^{-\beta\Delta_F}\coth(\beta\omega/2),
  \label{eq:Sparallel}
\end{align}
where $D=2JS$. We readily observe that the power spectrum is sub-ohmic, i.e.,
it diverges at low frequencies
$S_\parallel^{3D}(\omega)\propto1/\sqrt{\omega}$---this is a direct consequence
of the fact that longitudinal fluctuations are gapless. Due to this divergence,
the perturbation theory (Born approximation) cannot be used when there is
longitudinal coupling to the longitudinal noise. In
order to deal with this singularity, we study transverse ($Y$) and longitudinal ($X$)
coupling separately. The transverse coupling can be treated perturbatively,
while for the longitudinal coupling we solve the problem exactly.

\subsubsection{Transverse coupling to longitudinal noise}
The part of the Hamiltonian that describes transverse coupling to the longitudinal
noise reads
\begin{equation}
  \label{eq:trans2long}
  H=H_F+\sigma^{+}\color{black}\otimes \int d\bm r b_{\bf r}S_{\bf r}^{\tilde z}+\text{h.c.}
\end{equation}
Using Eq.~(\ref{eq:T1T2}) and the inequality $$S_\parallel^{\rm 3D}(\omega, {\bf
r})\le S_\parallel^{\rm 3D}(\omega, {\bf r}=0),$$ we obtain the relaxation time
\begin{align}
  T_1^{-1}&=\int d\bm r d\bm r^\prime b_{\bf r}b_{\bf r^\prime}S_\parallel^{\rm
  3D}(\Delta,{\bf r-r^\prime})\nonumber\\
  &\le\int d\bm r d\bm r^\prime b_{\bf r}b_{\bf r^\prime}S_\parallel^{\rm
  3D}(\Delta,{\bf r}=0)\nonumber\\
  &=b^2 S_\parallel^{\rm
  3D}(\Delta).
  \label{eq:T1trans2long}
\end{align}
The above expression readily shows that relaxation time can be tailored
arbitrarily small by choosing the ratio $T/\Delta_F$ sufficiently small.

\subsubsection{Longitudinal coupling to longitudinal noise}
Here we consider only longitudinal coupling to longitudinal noise thus the
Hamiltonian reads
\begin{equation}
  \label{eq:long2long}
  H=H_F+\sigma^{z}\otimes V\,,
\end{equation}
with $V=\int d\bm r a_{\bf r}S_{\bf r}^{\tilde z}$. To simplify the problem
further,~\cite{makhlin_dephasing_2004} we substitute $S_{\bf r}^{\tilde
z}\rightarrow S_{\bf r}^{\tilde x}$ since the latter is linear in magnon
operators while the former is quadratic. When the final formula for the
decoherence time is obtained we substitute back the power spectrum of $S_{\bf
r}^{\tilde z}$ instead of $S_{\bf r}^{\tilde x}$.
  
In order to study decoherence we have to calculate the following
quantity~\cite{makhlin_dephasing_2004}
\begin{align}
  \label{eq:msigma}
  \langle\sigma^{-}(t)\rangle &=e^{i\varepsilon t/\hbar} 
  \langle\sigma^{-}(0)\rangle\times\\
  &\times\left\langle
  \tilde T\!\exp\left(i\int\nolimits_{0}^{t} V dt'\right)
  T\!\exp\left(i\int\nolimits_{0}^{t} V dt'\right)
  \right\rangle\,,\nonumber
\end{align}
with ($\tilde T$) $T$ the (anti-) time ordering operator.
The average in the above expression can be evaluated using a cluster
expansion~\cite{abrikosov1975methods} and since the perturbation $V$ is linear
in the bosonic operators, only the second order cluster contributes. Therefore,
the final exact result for the time-evolution of $\sigma^-(t)$ reads

\begin{align}
 \langle\sigma^{-}(t)\rangle =e^{i\varepsilon t/\hbar}
 \langle\sigma^{-}(0)\rangle e^{-\frac{1}{2}\int_0^t\int_0^tS(t_2-t_1) dt_1dt_2 }\,,
  \label{eq:sigmamt}
\end{align}
where $S(t)=\langle[V(t),V(0)]_+\rangle$. After performing the Fourier transformation we
obtain

\begin{align}
  \label{eq:sigma_omega}
  \langle \sigma_{-}(t)\rangle &=e^{i\varepsilon t/\hbar}
  \langle\sigma_{-}(0)\rangle\times\\
  &\times\exp\left(
  -\frac{1}{2}\int\,\frac{d\omega}{2\pi} S(\omega)
  \frac{\sin^2(\omega t/2)}{(\omega/2)^2}
  \right)\,.\nonumber
\end{align}
Note that this expression is of  exactly the same form as the one for a
\textit{classical} Gaussian noise.~\cite{makhlin_dissipative_2004} Now we
substitute back $S_{\bf r}^{\tilde x}\rightarrow S_{\bf r}^{\tilde z}$
\begin{eqnarray}
&&\langle \sigma^{-}(t)\rangle=e^{i\varepsilon t/\hbar}
  \langle\sigma^{-}(0)\rangle\times\nonumber\\
 &&\times\exp\left(
  -\frac{1}{2}\int\,\frac{d\omega}{2\pi} \int d\bm r d\bm r^\prime a_{\bf
  r}a_{\bf r^\prime}S_\parallel^{\rm 3D}(\omega,{\bf r-r^\prime}) \frac{\sin^2(\omega
  t/2)}{(\omega/2)^2}
  \right)\,.\nonumber\\
\end{eqnarray}
 
For long times $t\gg\hbar/T$ the dynamics is of the form
\begin{equation}
  \langle \sigma^{-}(t)\rangle\sim
  e^{-2\sqrt{2\pi}a^2T^{5/2}e^{-\beta\Delta_F}t^{3/2}/(3D^3)+i\Delta t},
\end{equation}
where we have used the inequality $S_\parallel^{\rm 3D}(\omega, {\bf
r})\le S_\parallel^{\rm 3D}(\omega, {\bf r}=0)$. Thus, this type of decoherence
can be suppressed by choosing the ratio $T/\Delta_F$ sufficiently small.

\subsection{Transverse noise} 
The power spectrum of transverse fluctuations of the ferromagnet is gapped
and thus vanishes for $\omega<\Delta_F$ (see Appendix),
\begin{eqnarray}
 S_\perp^{\rm
  3D}(\omega)&=&0\,,\quad\hspace{3.35cm}\omega<\Delta_{F}\,,\\
  S_\perp^{\rm
  3D}(\omega)&=&\frac{S\sqrt{\omega-\Delta_F}}{D^{3/2}}\coth(\beta\omega/2),\quad
  \omega>\Delta_F.
  \label{eq:Sperp}
\end{eqnarray}
Since the transverse fluctuations are gapped and the precession frequency of
the qubits is below the gap, this noise source does not contribute in the second
order (Born approximation) because only virtual magnons can be excited. 
In this section we choose the quantization axes such that qubit splitting is along the
$z$-axis, while the ferromagnet is polarized along the $x$-axis (see Fig.~\ref{fig:setup_fig}), this is done solely
for simplicity and all the conclusions are also valid for the most general case.
The Hamiltonian of the coupled system is of the form Eq.~(\ref{eq:NoiseModel}) with
operators $X$ ($Y$) 
\begin{align}
  X=&\frac{i}{2}\int d\bm rc_{\bf r}(S^+_{\bf r}-S^-_{\bf r}),\\
  Y^+=&-\frac{i}{8}\int d\bm r (a_{\bf r}S^+_{\bf r}+ b_{\bf r}S_{\bf r}^-),
\end{align}
with $S^\pm_{\bf r}=S^y_{\bf r}\pm iS^z_{\bf r}$ and the definitions 
\begin{align}
  a_{\bf r}&=B_{\bf r}+3C_{\bf r}-6A_{\bf r},\\
  b_{\bf r}&=B_{\bf r}+3C_{\bf r}+6A_{\bf r},\\
  c_{\bf r}&=B_{\bf r}-3A_{\bf r}^{\prime\prime},
\end{align}
where $A_{\bm r}$, $B_{\bm r}$, $C_{\bm r}$ are given by
Eqs.~(\ref{eq:A})-(\ref{eq:B}). To proceed further we perform the SW
transformation on the Hamiltonian given by Eq.~(\ref{eq:NoiseModel}). We ignore
the Lamb and Stark shifts and obtain the effective Hamiltonian
\begin{align}
  H=H_{\rm F}+\frac{\Delta}{2}\sigma^z+\sigma^z\otimes \tilde X_2+\sigma^+\otimes
  \tilde Y_2^-+\sigma^-\otimes\tilde Y_2^+,
  \label{eq:Hqubitz2nd}
\end{align}
where
\begin{align}
  \tilde X_2&=X_2-\langle X_2\rangle,\\
  \tilde Y_2^\pm&=Y_2^\pm-\langle Y_2^\pm\rangle,
\end{align}
with the following notation
\begin{align}
  X_2&=4(Y^+_\Delta Y^-+Y^+Y_\Delta^-),\\
  Y_2^+&=2(Y^+_\Delta X-X_0Y^+),\\
  X_\omega&=\frac{i}{2}\int d\bm r\bm r^\prime\chi_\perp(\omega,\bm r-\bm
      r^\prime)c_{\bf r}(S^+_{\bf r^\prime}-S^-_{\bf r^\prime}),\\
  Y^+_\omega&=-\frac{i}{8}\int d\bm r \bm r^\prime\chi_\perp(\omega,\bm r-\bm
      r^\prime)(a_{\bf r}S^+_{\bf r^\prime}+ b_{\bf r}S_{\bf r^\prime}^-).
\end{align}

The model given by Eq.~(\ref{eq:NoiseModel}) yields the following expressions
for the relaxation and decoherence times
\begin{align}
  T_1^{-1}&=S_{\tilde Y_2^-}(\omega=\Delta),\\
  T_2^{-1}&=\frac{1}{2}T_1^{-1}+S_{\tilde X_2}(\omega=0).
  \label{eq:T1T24th}
\end{align}

After a lengthy calculation we obtain the following expressions for $T_{1}$ and $T_{2}$ (see Appendix for a detailed derivation)
\begin{align}
  T_1^{-1}\le&
  \frac{B^4S^2\Delta_F^2}{2D^3}\left(\frac{1}{\Delta_F}+\frac{1}{\Delta_F-\Delta}\right)^2
  f\left(\frac{\Delta}{\Delta_F},\beta\Delta_F\right),\\
  T_2^{-1}\le&\frac{B^4S^2\Delta_F^2}{4D^3}\left(\frac{1}{\Delta_F}+\frac{1}{\Delta_F-\Delta}\right)^2f\left(\frac{\Delta}{\Delta_F},\beta\Delta_F\right)+\nonumber\\
  &\frac{B^4S^2\Delta_F^2}{2D^3(\Delta_F-\Delta)^2}f\left(0,\beta\Delta_F\right),
  \label{eq:T1T2final1}
\end{align}
with the function $f(x,y)$ defined as follows
\begin{equation}
  f(x,y)=\int_{1+x}^\infty dz
  \frac{\sqrt{z-1}}{e^{yz}-1}\frac{\sqrt{z-x-1}}{e^{y(z-x)}-1}.
  \label{eq:fxy}
\end{equation}

It is important to note that $f(x,y)\propto e^{-y}$, i.e., we obtain, as before for the longitudinal noise, that the effect of transverse fluctuations can be
suppressed by choosing the temperature much smaller than the excitation gap of
the ferromagnet. As anticipated, Eq.~(\ref{eq:T1T2final1}) shows that
the transverse noise becomes more important as the resonance is approached
($\Delta\sim\Delta_F$).

\section{Estimates}
In this section we give numerical estimates for the coherent coupling mediated
by the ferromagnet and the associated decoherence times. These estimates are
valid for both silicon-based and NV-center qubits.

Assume that the qubits lie close to the disc axis at a distance $h=25\,nm$ below
the disc and that the ferromagnet has in-plane polarization (along $x$-axis).
Assume the thickness of the disk to be $20\,nm$, its radius to be $50\, nm$, and
a lattice constant of $4\angstrom$. In this case the stray field at the plane
$x=0$ is along $x$ and has a magnitude that can reach values up to $1\,T$
depending on the precise position of the qubit. Similarly, when the ferromagnet
is polarized out-of-plane (along the $z$-axis), then the stray field at position
$x=y=0$ is along $z$ and can take values up to $1\, T$. For these cases and when
the qubit splitting is brought close to resonance, $\Delta_{F}-\Delta\approx
10^{-2}\mu eV$, we obtain operation times on the order of tens of nanoseconds
when the qubits are separated by a distance of about $1\,\mu m$.  The
decoherence times $T_2$ depend strongly on the ratio $k_{B} T/\Delta_{F}$ and
the additional decoherence source can be made negligible if this ratio is
sufficiently small. For a magnon gap $\Delta_{F}=100\,\mu eV$ (corresponding to
a magnetic field of about $1\,T$) and a temperature $T=0.1\,K$, we obtain
decoherence times solely due to the coupling to the ferromagnet that are much
bigger than the operation times and the typical decoherence times of the qubits.

\section{Conclusions}
We propose a scheme to coherently couple two \textit{atomistic} qubits separated
over distances on the order of a micron. We present a sequence for the implementation of the entangling CNOT gate and obtain operation times on the order of a few tens of nanoseconds. We show that there is a regime where all fluctuations of the ferromagnet are under control and the induced decoherence is non-detrimental: this is achieved when the temperature is smaller than the excitation gap of the ferromagnet. The main novel aspect of our proposal is its applicability to the technologically very important silicon qubits and NV-centers to which previous coupling methods do not apply.
\section{Acknowledgements}
We would like to thank A.~Yacoby, A.~Morello, and R.~Warburton for useful
discussions. This work was supported by SNF, NCCR QSIT, and IARPA.

\appendix

\section{Holstein-Primakoff transformation}
For the sake of completeness we derive in this Appendix explicit expressions for
the different spin-spin correlators used in this work
\begin{equation}
C^{\alpha\beta}(\omega,{\bf q})=\langle S_{{\bf q}}^{\alpha}(\omega)S_{-{\bf q}}^{\beta}(0)\rangle\,.
\end{equation}

For this purpose, we make use of a Holstein-Primakoff
transformation
\begin{eqnarray}
S_{i}^{z}&=&-S+n_{i},\,\,\,S_{i}^{-}=\sqrt{2S}\sqrt{1-\frac{n_i}{2S}}a_i,\,\,\,\mathrm{and}\nonumber\\
S_{i}^{+}&=&\left(S_{i}^{-}\right)^{\dagger},
\end{eqnarray}
in the limit $n_{i}\ll 2S$, with $a_{i}$ satisfying bosonic commutation
relations and $n_{i}=a_{i}^{\dagger}a_{i}$.~\cite{nolting2009quantumtheory} The
creation operators $a_{i}^{\dagger}$ and annihilation operators $a_{i}$ satisfy
bosonic commutation relations and the associated particles are called magnons.
The corresponding Fourier transforms are straightforwardly defined as $a_{{\bf
q}}^{\dagger}=\frac{1}{\sqrt{N}}\sum_{i}e^{-i{\bf q}\cdot{\bf R}_{i}}a_{i}$. In
harmonic approximation, the Heisenberg Hamiltonian $H_{F}$ reads
\begin{equation}
H_{F}\approx\sum_{\bf q}\epsilon_{{\bf q}}a_{{\bf q}}^{\dagger}a_{{\bf q}}\,,
\end{equation}
where $\epsilon_{\bf q}=\omega_{\bf
q}+\Delta_F=4JS[3-(\cos(q_x)+\cos(q_y)+\cos(q_z))]+\Delta_{F}$ is the spectrum
for a cubic lattice with lattice constant $a=1$ and the gap $\Delta_{F}$ is
induced by the external magnetic field or anisotropy of the ferromagnet.

\section{Transverse correlators $\langle S_{{\bf q}}^{+}(t)S_{-{\bf q}}^{-}(0)\rangle$}
Let us now define the Fourier transforms in the harmonic approximation
\begin{eqnarray}
S_{{\bf q}}^{+}&=&\frac{1}{\sqrt{N}}\sum_{i}e^{-i{\bf q}{\bf r}_{i}}S_{i}^{+}=\frac{\sqrt{2S}}{\sqrt{N}}\sum_{i}e^{-i{\bf q}{\bf r}_{i}}a_i^{\dagger}=\sqrt{2S}a_{-{\bf q}}^{\dagger}\nonumber\,,\\
S_{-{\bf q}}^{-}&=&\frac{1}{\sqrt{N}}\sum_{i}e^{i{\bf q}{\bf r}_{i}}S_{i}^{-}=\frac{\sqrt{2S}}{\sqrt{N}}\sum_{i}e^{i{\bf q}{\bf r}_{i}}a_i=\sqrt{2S}a_{-{\bf q}}\,.
\end{eqnarray}
From this it directly follows that
\begin{align}
C^{+-}(t,{\bf q})&=\langle S_{{\bf q}}^{+}(t)S_{-{\bf
q}}^{-}(0)\rangle\nonumber\\
&=2S\langle
a_{-{\bf q}}^{\dagger}(t)a_{-{\bf q}}\rangle=2S e^{i\epsilon_{\bf q}t}n_{{\bf
q}}\,,
\end{align}
with $\epsilon_{\bf q}\approx D{\bf q}^2+\Delta_F$ in the harmonic
approximation.

The Fourier transform is then simply given by
\begin{eqnarray}
C^{+-}(\omega,{\bf q})&=&\frac{1}{\sqrt{2\pi}}\int_{-\infty}^{\infty}dt
e^{-i\omega t}C^{+-}(t,{\bf q})\nonumber\\
&=&\underbrace{\frac{1}{\sqrt{2\pi}}\int_{-\infty}^{\infty}dte^{i(\epsilon_{\bf
q}-\omega)t}}_{\sqrt{2\pi}\delta(\epsilon_{\bf q}-\omega)}2S n_{\bf
q}\nonumber\\
&=&\sqrt{2\pi}2S\delta(\epsilon_{\bf q}-\omega)\frac{1}{e^{\beta\omega}-1}\,.
\end{eqnarray}
The corresponding correlator in real space is then simply given by
($q:=\vert{\bf q}\vert$)
\begin{eqnarray}
C^{+-}(\omega,{\bf r})&=&\frac{1}{(2\pi)^{3/2}}\int d{\bf q}e^{i{\bf q}{\bf
r}}C^{+-}(\omega,{\bf q})\\
&=&\frac{\sqrt{2\pi}}{(2\pi)^{3/2}}2S\frac{1}{e^{\beta\omega}-1}\int d{\bf q}\delta(D{\bf q}^2+\Delta_F-\omega)e^{i{\bf q}{\bf r}}\nonumber\\
&=&\frac{2S}{e^{\beta\omega}-1}\int_{-1}^{1}\int_{0}^{\infty}dqdxq^2\delta(D q^2+\Delta_F-\omega)e^{iq rx}\nonumber\\
&=&\frac{4S}{r}\frac{1}{e^{\beta\omega}-1}\int_{0}^{\infty}dq q \delta(Dq^2+\Delta_F-\omega)\sin(qr)\,.\nonumber
\end{eqnarray}
Let us now perform the following substitution
\begin{equation}
y=Dq^2,
\end{equation}
which gives for $\omega>\Delta_F$
\begin{align}
C^{+-}(\omega,{\bf r})&=\frac{4S/r}{2D(e^{\beta\omega}-1)}\int_{0}^{\infty}dy\delta(y+\Delta_F-\omega)\times\nonumber\\
&\times\sin\left(\sqrt{\frac{y}{D}}r\right)\\
&=\frac{2S}{D}\frac{1}{e^{\beta\omega}-1}\frac{\sin(\sqrt{(\omega-\Delta_F)/D}r)}{r}\,\nonumber.
\end{align}
We remark  that
\begin{equation}
C^{+-}(\omega,{\bf r})=0,\quad\omega<\Delta_F.
\end{equation}
We note the diverging behavior of the above correlation function for $\Delta_F=0$
and $\omega\rightarrow0$, namely
\begin{eqnarray}
\frac{1}{e^{\beta\omega}-1}\frac{\sin\left(\sqrt{\frac{\omega}{D}}r\right)}{r}\rightarrow
 \frac{1}{\sqrt{D}\beta}\frac{1}{\sqrt{\omega}}\,.
\end{eqnarray}
Similarly, it is now easy to calculate the corresponding commutators and
anticommutators. Let us define
\begin{eqnarray}
S_{\perp}(t,{\bf q}):=\frac{1}{2}\{S_{{\bf q}}^{+}(t),S_{-{\bf q}}^{-}(0)\}\,.
\end{eqnarray}
It is then straightforward to show that
\begin{eqnarray}
S_{\perp}(t,{\bf q})&=&Se^{i\epsilon_{\bf q}t}(1+2n_{{\bf q}})\,,
\end{eqnarray}
and therefore
\begin{eqnarray}
S_{\perp}(\omega,{\bf q})&=&\frac{S}{\sqrt{2\pi}}\int_{-\infty}^{\infty}e^{i(\epsilon_{\bf q}-\omega)t}(1+2n_{\bf q})\nonumber\\
&=&S\sqrt{2\pi}\delta(\epsilon_{\bf q}-\omega)\left(1+2\frac{1}{e^{\beta\omega}-1}\right).
\end{eqnarray}
Following essentially the same steps as the one performed above, we obtain the
3D real space anticommutator for $\omega>\Delta_F$
\begin{eqnarray}
  S^\mathrm{3D}_{\perp}(\omega,{\bf q})&=&S\coth(\beta\omega/2)\times\\
&&\times \int_{-1}^{1}\int_{0}^{\infty}dxdq q^2 e^{iqrx}\delta(\epsilon_{{\bf q}}-\omega)\nonumber\\
&=&\frac{S}{D}\coth(\beta\omega/2)\frac{\sin(\sqrt{(\omega-\Delta_F)/D}r)}{r}\,.\nonumber\\
\end{eqnarray}
Let us now finally calculate the transverse susceptibility defined as
\begin{eqnarray}
\chi_\perp(t,{\bf q})&=&-i\theta(t)[S_{{\bf q}}^{+}(t),S_{-{\bf q}}^{-}(0)]\,.
\end{eqnarray}
As before, in the harmonic approximation, one finds
\begin{eqnarray}
\chi_\perp(t,{\bf q})&=&i\theta(t)2Se^{i\epsilon_{\bf q}t}\,.
\end{eqnarray}
In the frequency domain, we then have
\begin{align}
\chi_\perp(\omega,{\bf q})&=\frac{2iS}{\sqrt{2\pi}}\int_{0}^{\infty}dte^{i(\epsilon_{\bf
q}-\omega)t-\eta t}\\
&=-\frac{2S}{\sqrt{2\pi}}\frac{1}{\epsilon_{\bf
q}-\omega+i\eta}\nonumber\,,
\end{align}
and thus in the small ${\bf q}$ expansion
\begin{eqnarray}
\chi_\perp(\omega,{\bf q})&=&-\frac{2S}{\sqrt{2\pi}}\frac{1}{D{\bf q}^2+\Delta_F-\omega+i\eta}\,.
\end{eqnarray}
\begin{widetext}
In real space, for the three-dimensional case, we  obtain
\begin{eqnarray}
\chi^{\text {3D}}_\perp(\omega,{\bf r})&=&-\frac{2S}{\sqrt{2\pi}}\frac{2\pi}{(2\pi)^{3/2}}\int_{0}^{\infty}\int_{-1}^{1}dxdqq^2\frac{1}{D{\bf q}^2+\Delta_{F}-\omega+i\eta}e^{iqrx}\nonumber\\
&=&-\frac{4S}{\sqrt{2\pi}}\frac{2\pi}{(2\pi)^{3/2}}\frac{1}{r}\int_{0}^{\infty}dqq\frac{1}{Dq^2+\Delta_F-\omega+i\eta}\sin(qr)\,.
\end{eqnarray}
Making use of the 
Plemelj formula we obtain for $\omega>\Delta_F$
\begin{eqnarray}
\chi^{\text {3D}}_\perp(\omega,{\bf r})&=&
-\frac{2S}{\sqrt{2\pi}}\frac{2\pi}{(2\pi)^{3/2}}\frac{1}{r}\int_{-\infty}^{\infty}dq
q\frac{1}{Dq^2+\Delta_F-\omega+i\eta}\sin(qr)\nonumber\\
&=&-\frac{2S}{\sqrt{2\pi}}\frac{2\pi}{(2\pi)^{3/2}}\frac{1}{r}
P\int_{-\infty}^{\infty}dq
\frac{q}{Dq^2+\Delta_F-\omega}\sin(qr)+i\frac{2S}{\sqrt{2\pi}}\frac{2\pi^2}{(2\pi)^{3/2}}\frac{1}{r}\int_{-\infty}^{\infty}dq
q\delta(Dq^2+\Delta_F-\omega)\sin(qr)\nonumber\\
&=&-\frac{S}{D}\frac{\cos(r\sqrt{(\omega-\Delta_F)/D})}{r}+i\frac{S}{2D}\frac{\sin(\sqrt{(\omega-\Delta_F)/D}r)}{r}\,.
\end{eqnarray}
\end{widetext}
It is worth pointing out that the imaginary part of the susceptibility vanishes,
\begin{equation}
\chi^{\text {3D}}_{\perp}(\omega,{\bf r})''=0,\quad \omega<\Delta_F,
\end{equation}
and therefore the susceptibility is purely real and takes the form of a Yukawa
potential
\begin{equation}
\chi_{\perp}^{\text{3D}}(\omega,{\bf r})=-\frac{S}{D}\frac{e^{-r/l_F}}{r},\quad\omega<\Delta_F,
\end{equation}
where $l_F=\sqrt{\frac{D}{\Delta_F-\omega}}$.

Note also that the imaginary part of the transverse susceptibility satisfies the
well-know fluctuation-dissipation theorem
\begin{eqnarray}
  S^\mathrm{3D}_{\perp}(\omega,{\bf r})&=&\coth(\beta\omega/2)\chi^{\text {3D}}_\perp(\omega,{\bf r})^{\prime\prime}\,.
\end{eqnarray}

In three dimensions the susceptibility decay as $1/r$, where $r$ is measured in
lattice constants. For distances of order of $1\mu m$ this leads to four orders
of magnitude reduction.

For quasi one-dimensional ferromagnets such a reduction is absent and the
transverse susceptibility reads
\begin{equation}
  \chi_\perp^{\text{1D}}(\omega,r)=-\frac{S}{D}l_F e^{-r/l_F},\quad\omega<\Delta_F,
  \label{eq:chi_1D_r}
\end{equation}
where $l_F$ is defined as above and the imaginary part vanishes as above, i.e.,
\begin{equation}
\chi_{\perp}^{\text{1D}}(\omega,r)''=0,\quad\omega<\Delta_{F}.
\end{equation}
Similarly for $\omega>\Delta_F$ we have
\begin{equation}
\chi_{\perp}^{\text{1D}}(\omega,r)=S\frac{\sin\left(\sqrt{(\omega-\Delta_F)/D}r\right)}{\sqrt{D(\omega-\Delta_F)}}\,,
\end{equation}
and
\begin{equation}
\chi_{\perp}^{\text{1D}}(\omega,r)''=\frac{S}{2D}\sqrt{\frac{D}{\omega-\Delta_F}}\cos\left(\sqrt{(\omega-\Delta_F)/D}r\right)\,.
\label{eq:chi_perp1D}
\end{equation}

\section{Longitudinal correlators $\langle S_{{\bf q}}^{z}(t)S_{-{\bf q}}^{z}(0)\rangle$}
The longitudinal susceptibility reads
\begin{eqnarray}
\chi_\parallel(t,{\bf q})&=&-i\theta(t)[ S_{{\bf q}}^{z}(t),S_{-{\bf
q}}^{z}(0)]\\ &=&-\theta(t)\frac{1}{N}\sum_{{\bf q}',{\bf
q}''}e^{it(\epsilon_{{\bf q}'}-\epsilon_{{\bf q}'+{\bf q}})}\langle[a_{{\bf
q}'}^{\dagger}a_{{\bf q}'+{\bf q}},a_{{\bf q}''}^{\dagger}a_{{\bf q}''-{\bf
q}}]\rangle\nonumber\,.
\end{eqnarray}
Applying Wick's theorem and performing a Fourier transform, we obtain the
susceptibility in frequency domain
\begin{equation}
\chi_\parallel(\omega,{\bf q})=-\frac{1}{N}\sum_{{\bf k}}\frac{n_{\bf k}-n_{{\bf k}+{\bf q}}}{\omega-\epsilon_{{\bf k}+{\bf q}}+\epsilon_{\bf k}+i\eta}\,,
\end{equation}
where $n_{\bf k}$ is the magnon occupation number given by the
Bose-Einstein distribution 
\begin{equation}
n_k=\frac{1}{e^{\beta \epsilon_{\bf k}}-1}\,,
\end{equation}
where $\epsilon_{\bf k}$ is again the magnon spectrum ($\epsilon_{\bf
k}=\omega_{\bf k}+\Delta_F\approx D{\bf k}^2+\Delta_F$ for small $k$). Note that
the longitudinal susceptibility is proportional to $1/S$, due to the fact that
$\epsilon_{\bf k}-\epsilon_{{\bf k}+{\bf q}}=\omega_{\bf k}-\omega_{{\bf k}+{\bf
q}}\propto S$.
 
Since we are interested in the decoherence processes caused by the longitudinal
fluctuations, we calculate the imaginary part of $\chi_\parallel(\omega,{\bf
q})$ which is related to the fluctuations via the
fluctuation-dissipation theorem. Performing a small ${\bf q}$
expansion and assuming without loss of generality $\omega>0$, we obtain for the imaginary part
\begin{widetext}
\begin{eqnarray}
\chi^{\text{3D}}_\parallel(\omega,{\bf q})^{\prime\prime}&=&
\frac{\pi}{(2\pi)^3}\int d{\bf
k}(n_{\bf k}-n_{{\bf k}+{\bf q}})\delta(\omega_{\bf k}-\omega_{{\bf k}+{\bf
q}}+\omega)\nonumber\\
&=&\frac{1}{4\pi}\int_{0}^{\infty}dk
k^2\int_{-1}^{1}dx\left(\frac{1}{e^{\beta(\Delta_F+ D k^2)}-1}-\frac{1}{e^{\beta
(\omega+\Delta_F+Dk^2)}-1}\right)\delta(\omega-D q^2-2Dkqx)\nonumber\\
&=&\frac{1}{4\pi}\int_{0}^{\infty}dk
k^2\int_{-1}^{1}dx\left(\frac{1}{e^{\beta(\Delta_F+D k^2)}-1}-\frac{1}{e^{\beta
(\omega+\Delta_F+Dk^2)}-1}\right)\delta\left(k-\frac{\omega-Dq^2}{2Dqx}\right)\left\vert\frac{1}{2Dqx}\right\vert\nonumber\\
&=&\frac{1}{4\pi}
\int_{-1}^{1}dx\left\vert\frac{1}{2Dqx}\right\vert\left(\frac{\omega-Dq^2}{2Dqx}\right)^2
\left(\frac{1}{e^{\beta
  \left(\Delta_F+D\left(\frac{\omega-Dq^2}{2Dqx}\right)^2\right)}-1}-\frac{1}{e^{\beta
    \left(\omega+\Delta_F+D\left(\frac{\omega-Dq^2}{2Dqx}\right)^2\right)}-1}\right)\theta\left(\frac{\omega-Dq^2}{2Dqx}\right)\nonumber\\
    &=&\frac{1}{4\pi}\int_{0}^{1}dx\frac{1}{2Dqx}\left(\frac{\omega-Dq^2}{2Dqx}\right)^2
\left(\frac{1}{e^{\beta
  \left(\Delta_F+D\left(\frac{\omega-Dq^2}{2Dqx}\right)^2\right)}-1}-\frac{1}{e^{\beta
    \left(\omega+\Delta_F+D\left(\frac{\omega-Dq^2}{2Dqx}\right)^2\right)}-1}\right).
    \label{eq:chizzq}
  \end{eqnarray}
\end{widetext}
Next, since we are interested in the regime where $\omega\gg T$ (and thus
$\beta\omega\gg 1$), we have $n_{\bm k}\gg n_{\bm{k+q}}$. Furthermore, we approximate
the distribution function $n_{\bm k}=\frac{e^{-\beta(\Delta_F+\omega_{\bm
k})}}{1-e^{-\beta\Delta_F}+\beta\omega_{\bm k}}$ (this is valid when
$\beta\omega_{\bf k}\ll 1$) and arrive at the following expression
\begin{widetext}
\begin{align}
    \label{eq:qzz}
    \chi^{\text{3D}}_\parallel(\omega,{\bf q})^{\prime\prime}&=\frac{1}{4\pi}\int_{0}^{1}dx\frac{1}{2Dqx}\left(
    \frac{\omega-Dq^2}{2Dqx} \right)^2 \frac{e^{-\beta\left(\Delta_F+D\left(
      \frac{\omega-Dq^2}{2Dqx} \right)^2\right)}}{1-e^{-\beta\Delta_F}+\beta D\left(
    \frac{\omega-Dq^2}{2Dqx} \right)^2 }\nonumber\\
    &=-\frac{e^{1-e^{-\beta\Delta_F}-\beta\Delta_F}}{4\beta D^2q}\mathrm {Ei}\left(e^{-\beta\Delta_F}+\frac{1}{4}\left(-4-\beta
    Dq^2+2\beta\omega-\frac{\beta\omega^2}{Dq^2}\right)\right)\,,
\end{align}
\end{widetext}
where $\mathrm{Ei}(z)$ is the exponential integral function. We also need the
the real space representation obtained after inverse Fourier transformation,
\begin{align}
  \chi^{\text{3D}}_\parallel(\omega,{\bf r})^{\prime\prime}&=\sqrt{\frac{2}{\pi}}\frac{1}{r}\int_{0}^{\infty}dqq
\chi^{\text{3D}}_\parallel(\omega,q)^{\prime\prime} \sin(qr)\,.
\label{eq:chi_zzr}
\end{align}

In order to perform the above integral we note that the imaginary part of the
longitudinal susceptibility, given by Eq.~(\ref{eq:qzz}), is peaked
around $q=\sqrt{\omega/D}$ with the width of the peak ($1/\sqrt{\beta D}$)
much smaller than its position in the regime we are working in ($\omega\gg T$).
For $\bf r=0$, the integration over $q$ can be then performed approximately and
yields the following expression

\begin{align}
  \chi^{\text{3D}}_\parallel(\omega,{\bf r=0})^{\prime\prime}&=\frac{\sqrt{\pi}e^{-e^{-\beta\Delta_F}-3\beta\Delta_F/2}}{2\beta^2D^3}
  \left( e^{e^{-\beta\Delta_F}+\beta\Delta_F/2}\right.\nonumber\\
  &-e\sqrt{\pi}\sqrt{e^{\beta\Delta_F}-1}\\
  &\times\left.\mathrm{Erfc}(e^{-\beta\Delta_F/2}\sqrt{e^{\beta\Delta_F}-1})\right)\sqrt{\beta\omega}\,,\nonumber
\end{align}
where $\mathrm{Erfc}(z)$ denotes the complementary error function. It is
readily observed from the above expression that the longitudinal fluctuations
are exponentially suppressed by the gap. Assuming that $\Delta_F\gg T$, we
obtain the following simplified expression
\begin{align}
  \chi^{\text{3D}}_\parallel(\omega,{\bf r=0})^{\prime\prime}&=\frac{\sqrt{\pi}-e\pi\mathrm{Erfc}(1)}{2\beta^2D^3}e^{-\beta\Delta_F}\sqrt{\beta\omega}\,.
  \label{eq:spectraldensity_longitudinal}
\end{align}
We observe that, since $J(\omega)=\chi_\parallel(\omega,{\bf
r})^{\prime\prime}$, the longitudinal noise of the ferromagnet is---as the
transverse one---sub-ohmic.~\cite{divincenzo_rigorous_2005}

It is interesting to obtain the behavior of the longitudinal susceptibility in
the opposite limit, when $\beta\omega\ll1$. In this limit, the difference of the
two Boltzmann factors in Eq.~(\ref{eq:chizzq}) can be expanded to the lowest
order in the small quantity $\beta\omega$,
\begin{align}
  \chi_\parallel^{\rm 3D}(\omega,\bm q)^{\prime\prime}=&\int_0^1dx\frac{1}{8\pi
  Dqx}\left(\frac{\omega-Dq^2}{2Dqx}\right)^2\times\nonumber\\
  &\frac{\beta\omega}{\ch\left(\beta\Delta_F+\beta
  D\left(\frac{\omega-Dq^2}{2Dqx}\right)^2
  \right)-1}\nonumber\\
  =&\frac{\omega}{16\pi D^2q\left(
  e^{\beta\Delta_F+\frac{\beta(\omega-Dq^2)^2}{4Dq^2}}-1 \right)}.
  \label{eq:chi_x_p}
\end{align}
In order to calculate the Fourier transform to real space, we note that for
$\beta\omega\ll1$ the denominator of the above expression depends only weakly on
$\omega$, thus we ignore this dependence and obtain the Fourier transform for
$\bm r=0$
\begin{align}
  \chi_\parallel^{\rm 3D}(\omega)^{\prime\prime}=\frac{\ln(1+n_{\bm
  k=0})}{16\pi\beta D^3}\omega.
  \label{eq:chizz_imag_highT}
\end{align}
The above formula shows that the longitudinal noise of a ferromagnet at high
temperatures ($\beta\omega\ll1$) behaves as ohmic rather than sub-ohmic bath.

Next we calculate the longitudinal fluctuations for the case of a quasi-
one-dimensional ferromagnet ($\Delta_F\gg T$) and obtain

\begin{align}
  \chi_\parallel^{\text{1D}}(\omega,r=0)^{\prime\prime}&=\frac{1}{4\pi}\int_{-\infty}^{\infty}dk\int_{-\infty}^{\infty}dq
\left(\frac{1}{e^{\beta(\Delta_F+ D k^2)}-1}\right.\nonumber\\
&-\left.\frac{1}{e^{\beta (\omega+\Delta_F+Dk^2)}-1}\right)\delta(\omega-D
q^2-2Dkq)\nonumber\\
&=\int_{-\infty}^\infty dk \frac{e^{-\beta Dk^2}}{1-e^{-\beta\Delta_F}+\beta
Dk^2}\frac{1}{D\sqrt{k^2+\omega/D}}\nonumber\\
&=\frac{\gamma}{D\sqrt{\beta\omega}}e^{-\beta\Delta_F}\,,
  \label{eq:fluct_zz_1D}
\end{align}
where $\gamma$ is a numerical factor of order unity. 

Note that $S_{\parallel}(\omega,{\bf r})$ is defined through the fluctuation dissipation theorem as
\begin{equation}
S_{\parallel}(\omega,{\bf r})=\coth(\beta\omega/2)\chi_{\parallel}(\omega,{\bf r})^{\prime\prime}\,.
\end{equation}

\section{Exchange coupling to the ferromagnet}
\subsubsection{Exchange coupling}
The Hamiltonian we consider is of the following form
\begin{equation}\label{eq:Hamiltonian}
H=H_F+H_\sigma+A\sum_{\bm i}\bm\sigma_i\cdot\bm S_{{\bf r}_i}\,,
\end{equation}
where $A$ is the exchange coupling constant between the qubit spins and the
ferromagnet. 
The ferromagnet is assumed to be below the Curie
temperature with the magnetization pointing along the out-of-plane $z$-direction. The qubit
Hamiltonian is assumed to be without splitting initially, that is
$H^{(0)}_\sigma=0$. Nevertheless, since the ferromagnet is in the ordered phase,
there exists a first order effect due to coupling to the ferromagnet which gives
rise to the term of the form $A\sum_i\sigma^z_i\langle S^z_{{\bf r}_i}\rangle$.
Such a splitting is undesirable if one is interested
in \textit{coherent} interaction---we remedy this by coupling the spins to
another ferromagnet, albeit with anti-parallel magnetization~\footnote{Note that
the $z$-component of the magnetization of both the ferromagnets does not need to
cancel exactly. We only require the splitting along $z$ to be smaller that
$\Delta_F$.}. Since we allow for some misalignment between orientation of the
magnetization of the two ferromagnets, the final Hamiltonian for the qubits in
the spin space after taking into account the first order corrections due to
coupling to the ferromagnet reads
\begin{equation}
  H_\sigma=\frac{1}{2}\Delta\sum_i \sigma^x_i\,.
  \label{eq:1stH}
\end{equation}
The splitting in the $x$-direction of the qubit (or equivalently along the
$y$-direction) is beneficial since it reduces decoherence due to longitudinal
noise of the ferromagnet: the effect of such noise spectrum can significantly
influence decoherence times for the case of no splitting of the qubit because the
longitudinal noise is gapless.

\subsubsection{Coherent coupling}
We proceed with the derivation of an effective two-spin interaction Hamiltonian for
$A\ll J$ by employing a perturbative Schrieffer-Wolff
transformation~\cite{bravyi_schrieffer_2011} up to the second order
\begin{align}
H_{\text{eff}}&=H_\sigma+\frac{A^2}{8}\chi_{\perp}(\Delta)(2\sigma_{1}^y\sigma_{2}^y+\sigma_{1}^{z}\sigma_{2}^x+\sigma_{1}^x\sigma_{2}^z)\,,
\label{eq:Heff_supp}
\end{align}
where we introduced the notation $\chi_\perp(\omega)=\chi_\perp(\omega,L)$
($L=\vert {\bf r}_2-{\bf r}_1\vert$) and $\chi_\perp(\omega,{\bf r})$ is the
transverse real space spin susceptibility of the ferromagnet. Note that we have
neglected $\chi^{\text{3D}}_\perp(-\Delta)$ and $\chi^{\text{3D}}_\perp(0)$ in
comparison to $\chi^{\text{1D}}_\perp(\Delta)$, as well as the
longitudinal susceptibility $\chi_\parallel$ since it is smaller by factor of
$1/S$ compared to the transverse one and it is suppressed by temperature. The real space transverse
susceptibility of the 3D ferromagnet is given by
\begin{equation}
  \chi^\mathrm{3D}_\perp(\omega,{\bf r}) =-\frac{S}{D}
  \frac{e^{-r/l_F}}{r},\quad \omega<\Delta_F,
   \label{eq:chi_r}
\end{equation}
where $\Delta_F$ is the gap induced via applied external magnetic field or due
to internal anisotropy of the ferromagnet,
$l_F=\sqrt{\frac{D}{\Delta_F-\omega}}$ and $D=2JS$. In what follows, we assume
that the external gap is always larger than the qubit splitting,
$\Delta<\Delta_F$, as this ensures that the transverse noise is not
contributing to decoherence in second order since transverse noise is related to the vanishing
imaginary part of the transverse susceptibility,
$\chi_\perp(\omega)^{\prime\prime}=0$ ($\omega<\Delta_F$). The spatial
dependence of the effective two spin coupling given by Eq.~(\ref{eq:chi_r}) is
of Yukawa type due to presence of the external gap. If we assume a realistic
tunnel coupling to the ferromagnet of $100\mu
\mathrm{eV}$,~\cite{hauptmann2008electricfieldcontrolledspin,hofstetter_ferromagnetic_2010}
the Curie temperature of $550\mathrm K$ [as for example for yttrium iron garnet
(YIG)] and a gap of $\Delta_F=100\mu\mathrm{eV}$,  and the qubit splitting
close to the resonance $\Delta_F-\Delta=3\times 10^{-3}\mu \mathrm{eV}$
(corresponding to a magnetic field of about $B=60\mu\mathrm{T}$) we obtain for
the qubit-qubit coupling strength a value on the order of
$4\times10^{-11}\,\mathrm{eV}$ for a lattice constant of about $4\angstrom$.
This coupling strength gives rise to the operation times of $5\mu
s$---significantly below the relaxation and decoherence times of the spin
qubit, $T_1=1s$~\cite{amasha_electrical_2008} and $T_2>200\mu
s$~\cite{Bluhm_Dephasing_2011} respectively. Furthermore, the error
threshold---defined as the ratio between the two-qubit gate operation time to
the decoherence time---we obtain with such an operation time is about
$10^{-2}$, which is good enough for implementing the surface code error
correction.~\cite{raussendorf_fault-tolerant_2007} Here we used $T_2$ instead
of $T_2^*$ since spin-echo can be performed together with two-qubit
gates.~\cite{khodjasteh_dynamically_2009} Alternatively, the decoherence time
of GaAs qubits can be increased without spin-echo by narrowing the state of the
nuclear spins.~\cite{Xu_Optically_2009,Vink_Locking_2009}

The dimensionality of the ferromagnet plays an important role---if we assume
$10nm$ width of the trench where the ferromagnet is placed, then, for  energies
below $0.1m \mathrm{eV}$, the ferromagnet behaves as quasi one-dimensional
(1D). In this case we obtain
\begin{equation}
  \chi_\perp^{\text{1D}}(\omega,r)=-\frac{S}{D} l_F
  e^{-r/l_F},\quad\omega<\Delta_F,
  \label{eq:chi1D}
\end{equation}
wherefrom it is evident that at distances $r\lesssim l_F$ the susceptibility of
a quasi-1D ferromagnet is practically constant in contrast to the 3D case,
where a $1/r$ decay is obtained, see Eq.~(\ref{eq:chi_r}). Additionally, we
require $l_F\lesssim D/(AS)=2J/A$ for the perturbation theory to be valid.
Thus, for the same parameters as above, but without the need to tune very close
to the resonance (we set herein $\Delta_F-\Delta=0.5\mu \mathrm{eV}$,
corresponding to about $B=10 m\mathrm{T}$) a coupling strength of
$10^{-8}\mathrm{eV}$ is obtained.

For 1D case there is yet another rather promising possibility---to use magnetic
semiconductors.~\cite{ohno2000electricfieldcontrol} These materials are
characterized by a particularly low Curie temperature of $30$K or
below,~\cite{ohno2000electricfieldcontrol} and the distance between the ions
that are magnetically ordered via RKKY interaction is about $10-100nm$. Such a
large lattice constant is very beneficial for the long range coupling---if we
take the lattice constant to be $10nm$, the coupling to the ferromagnet
$A=15\mu \mathrm{eV}$ and the qubit splitting close to resonance
($\Delta_F-\Delta=0.5\mu \mathrm{eV}$, corresponding to about
$B=10m\mathrm{T}$), the qubit-qubit coupling becomes of the order of $1\mu
\mathrm{eV}$. Such a coupling strength in turn leads to an error threshold on
the order of $10^{-8}$. Therefore, even the standard error correction protocol
can be used in this case.
\subsubsection{Derivation of the effective Hamiltonian (exchange coupling)}
Here we give a detailed derivation of the qubit-qubit effective Hamiltonian. As stated above, the total Hamiltonian of the system reads
\begin{equation}
H=H_F+H_\sigma+A\sum_{\bm i}\left(\frac{1}{2}(\sigma_{i}^{+}S_{{\bf r}_i}^{-}+\sigma_{i}^{-}S_{{\bf r}_i}^{+})+\sigma_{i}^zS_{{\bf r}_i}^z\right)\,,
\end{equation}
where we identify the main part as $H_{0}=H_F+H_\sigma$ and the small
perturbation as the exchange coupling $V=A\sum_{\bm i}\bm\sigma_i\cdot\bm
S_{\bm r_i}$. The Hamiltonian of the ferromagnet  reads $H_{F}=-J\sum_{\langle
\bm r,\bm r'\rangle}{\bm S}_{\bm r}\cdot{\bm S}_{\bm r'}$, while the
Hamiltonian for the two distant qubits is $
H_\sigma=\frac{\Delta}{2}\sum_{i=1,2} \sigma^x_i$.

The second order effective Hamiltonian~\cite{bravyi_schrieffer_2011} is given
by $H_{\text{eff}}^{(2)}=H_{0}+U$, where
\begin{equation}\label{eq:SW}
U=-\frac{i}{2}\lim\limits_{\eta\rightarrow 0^{+}}\int_{0}^{\infty}dte^{-\eta t}[V(t),V]\,,
\end{equation}
where $V(t)=e^{i H_{0}t}Ve^{-iH_{0}t}$.

We have
\begin{equation}\label{eq:time_evolution}
\sigma_{i}^{+}(t)=\frac{1+\cos(\Delta t)}{2}\sigma_{i}^{+}+\frac{1-\cos(\Delta t)}{2}\sigma_{i}^{-}-i\sin(\Delta t)\sigma_{i}^z\,,
\end{equation}
and $\sigma_{i}^{-}(t)=\sigma_{i}^{+}(t)^{\dagger}$.

Recalling that the $zz$ susceptibility can be neglected and that only the
transverse susceptibility contributes, we obtain the following result from
Eq.~(\ref{eq:SW}), $U=\lim\limits_{\eta\rightarrow0^{+}}\int_{0}^{\infty}dt
e^{-\eta t}\sum_{ij}U_{ij}$
\begin{align}
\label{eq:U2}
U_{ij}=&-\frac{iA^2}{8}\left([\sigma_{i}^{-}(t)S_{{\bf
r}_i}^{+}(t),\sigma_{j}^{+}S_{{\bf r}_j}^{-}]+\text{h.c.}\right)\nonumber\\
=&-\frac{iA^2}{8}\left(\sigma_{i}^{-}(t)\sigma_{j}^{+}[S_{{\bf
r}_i}^{+}(t),S_{{\bf r}_j}^{-}]+\text{h.c.}\right)
\end{align}
Finally, by rewriting $\cos(\Delta t)=\frac{e^{i\Delta t}+e^{-i\Delta t}}{2}$,
$\sin(\Delta t)=\frac{e^{i\Delta t}-e^{-i\Delta t}}{2i}$, and using the
definition of the real space transverse spin susceptibility
\begin{equation}
\chi_{\perp}(\omega,{\bf r}_{i}-{\bf
r}_{j})=-i\lim\limits_{\eta\rightarrow0^{+}}\int_{0}^{\infty}dt e^{-(i\omega+\eta) t}[S_{{\bf r}_i}^{+}(t),S_{{\bf r}_j}^{-}]\,,
\end{equation}
we obtain by inserting Eq.~(\ref{eq:time_evolution}) into Eq.~(\ref{eq:U2})
\begin{align}
    U=&\frac{A^2}{8}\sum_{ij}\left(\frac{\chi_{\perp}(0)}{2}+\frac{\chi_{\perp}(\Delta)+\chi_{\perp}(-\Delta)}{4}\right)\sigma_{i}^{-}\sigma_{j}^{+}\nonumber\\
    +&\frac{A^2}{8}\sum_{ij}\left(\frac{\chi_{\perp}(0)}{2}-\frac{\chi_{\perp}(\Delta)+\chi_{\perp}(-\Delta)}{4}\right)\sigma_{i}^{+}\sigma_{j}^{+}\nonumber\\
    -&\frac{A^2}{8}\sum_{ij}\frac{\chi_{\perp}(\Delta)-\chi_{\perp}(-\Delta)}{2}\sigma_{i}^{z}\sigma_{j}^{+}+\text{h.c.}
\end{align}
Since the decay length of the susceptibility $\chi(\omega,{\bf r})$ is large
only close to the resonance, $\Delta_F\sim\Delta$, we can simplify the above
equation by neglecting $\chi(-\Delta,{\bf r})$ and $\chi(0,{\bf r})$ in
comparison to $\chi(\Delta,{\bf r})$ which is assumed to be close to the
resonance. Within this approximation we arrive at Eq.~(\ref{eq:Heff_supp}) of
the main text.

\section{Fourth order contributions to decoherence}
In this section we determine the effect of the transverse noise in the lowest
non-vanishing order due to coupling dipolarly to the ferromagnet. Here we
choose quantizations axes such that the qubit splitting is along the $z$-axis,
while the ferromagnet is polarized along $x$-axis. The Hamiltonian of the
coupled system reads
\begin{align}
  H=H_{\rm F}+\frac{\Delta}{2}\sigma^z+\sigma^z\otimes X+\sigma^+\otimes
  Y^-+\sigma^-\otimes Y^+,
  \label{eq:Hqubitz_supp}
\end{align}
where the operator $X$ ($Y$) that couples longitudinally (transversally) to the
qubit is linear in the transverse operators of the ferromagnet
\begin{align}
  X=&\frac{i}{2}\int d\bm rc_{\bf r}(S^+_{\bf r}-S^-_{\bf r}),\\
  Y^+=&-\frac{i}{8}\int d\bm r (a_{\bf r}S^+_{\bf r}+ b_{\bf r}S_{\bf r}^-),
\end{align}
with $S^\pm_{\bf r}=S^y_{\bf r}\pm iS^z_{\bf r}$ and the definitions of the coefficients
\begin{align}
  a_{\bf r}&=B_{\bf r}+3C_{\bf r}-6A_{\bf r},\\
  b_{\bf r}&=B_{\bf r}+3C_{\bf r}+6A_{\bf r},\\
  c_{\bf r}&=B_{\bf r}-3A_{\bf r}^{\prime\prime},\\
  A_{\bf r}&=\frac{1}{a^{3}}\frac{r^{z}r^{+}}{r^5}\,,\label{eq:A_supp}\\
  C_{\bf r}&=\frac{1}{a^{3}}\frac{(r^+)^2}{r^5}\,,\label{eq:C_supp}\\
  B_{\bf r}&=\frac{1}{a^{3}}\frac{1}{r^3}\left(2-\frac{3r^+r^-}{r^2}\right)\,\label{eq:B_supp}\,.
\end{align}
To proceed further we perform the SW transformation on the Hamiltonian given by
Eq.~(\ref{eq:Hqubitz_supp}). We ignore the Lamb and Stark shifts and obtain the
effective Hamiltonian
\begin{align}
  H=H_{\rm F}+\frac{\Delta}{2}\sigma^z+\sigma^z\otimes \tilde X_2+\sigma^+\otimes
  \tilde Y_2^-+\sigma^-\otimes\tilde Y_2^+,
  \label{eq:Hqubitz2nd_supp}
\end{align}
where
\begin{align}
  \tilde X_2&=X_2-\langle X_2\rangle,\\
  \tilde Y_2^\pm&=Y_2^\pm-\langle Y_2^\pm\rangle,
\end{align}
with the following notation
\begin{align}
  X_2&=4(Y^+_\Delta Y^-+Y^+Y_\Delta^-),\\
  Y_2^+&=2(Y^+_\Delta X-X_0Y^+),\\
  X_\omega&=\frac{i}{2}\int d\bm r\bm r^\prime\chi_\perp(\omega,\bm r-\bm
      r^\prime)c_{\bf r}(S^+_{\bf r^\prime}-S^-_{\bf r^\prime}),\\
  Y^+_\omega&=-\frac{i}{8}\int d\bm r \bm r^\prime\chi_\perp(\omega,\bm r-\bm
      r^\prime)(a_{\bf r}S^+_{\bf r^\prime}+ b_{\bf r}S_{\bf r^\prime}^-),
\end{align}

The model given by Eq.~(\ref{eq:Hqubitz2nd_supp}) yields the following expressions
for the relaxation and decoherence times
\begin{align}
  T_1^{-1}&=S_{\tilde Y_2^-}(\omega=\Delta),\\
  T_2^{-1}&=\frac{1}{2}T_1^{-1}+S_{\tilde X_2}(\omega=0),
  \label{eq:T1T24th_supp}
\end{align}
where, again, $S_A(\omega)=\int dt e^{-i\omega
t}\{A^\dagger(t),A(0)\}$.

After a lengthly calculation we obtain the expressions for $S_{\tilde
X_2}(\omega=0)$ and $S_{\tilde Y_2^-}(\omega=\Delta)$,
\begin{widetext}
  \begin{align}
    S_{\tilde X_2}(0)=&\frac{1}{128} \int d\nu d\bm r_1\bm r_2\bm r_3\bm
    r_4\bm r_5\bm r_6C^{-+}(\nu ,\bm r_3-\bm
    r_4) C^{+-}(-\nu ,\bm r_1-\bm r_2)\times\\
    (&(a_{{\bf r}_5} a^*_{{\bf r}_3}+b_{{\bf r}_3} b^*_{{\bf r}_1}) (a_{{\bf r}_4}
    a^*_{{\bf r}_2}+b_{{\bf r}_6} b^*_{{\bf r}_4}) \chi _\perp(\Delta ,\bm r_1-\bm
    r_5) \chi _\perp(\Delta ,\bm r_2-\bm r_6)+\nonumber\\
    &(a_{{\bf r}_4} a^*_{{\bf r}_2}+b_{{\bf r}_5} b^*_{{\bf r}_4}) (a_{{\bf r}_1} a^*_{{\bf r}_3}+b_{{\bf r}_6}
    b^*_{{\bf r}_1}) \chi _\perp(\Delta ,\bm r_2-\bm r_5) \chi
    _\perp(\Delta ,\bm r_3-\bm r_6)+\nonumber\\
    &(a_{{\bf r}_6} a^*_{{\bf r}_2}+b_{{\bf r}_2} b^*_{{\bf r}_4}) (a_{{\bf r}_5} a^*_{{\bf r}_3}+b_{{\bf r}_3}
    b^*_{{\bf r}_1}) \chi _\perp(\Delta ,\bm r_1-\bm r_5) \chi
    _\perp(\Delta ,\bm r_4-\bm r_6)+\nonumber\\
    &(a_{{\bf r}_6} a^*_{{\bf r}_2}+b_{{\bf r}_2} b^*_{{\bf
    r}_4}) (a_{{\bf r}_1} a^*_{{\bf r}_3}+b_{{\bf r}_5} b^*_{{\bf
    r}_1}) \chi _\perp(\Delta ,\bm r_3-\bm r_5) \chi
    _\perp(\Delta ,\bm r_4-\bm r_6))\, ,\nonumber\\
    \label{eq:SX2_supp}
    S_{\tilde Y^-_2}(\Delta)=&\frac{1}{64} \int d\nu d\bm r_1\bm r_2\bm r_3\bm
        r_4\bm r_5\bm r_6C^{-+}(\nu ,\bm r_3-\bm r_4)
        C^{+-}(\Delta-\nu ,\bm r_1-\bm r_2)\times\\
        (&c_{{\bf r}_3} c_{{\bf r}_6} (a_{{\bf r}_4} b_{{\bf r}_1}^*+a_{{\bf
        r}_5} b_{{\bf r}_4}^*) \chi_\perp(0,\bm r_2-\bm r_6)
        \chi_\perp(\Delta ,\bm r_1-\bm r_5)-c_{{\bf r}_3} c_{{\bf r}_6} (a_{{\bf r}_5} a_{{\bf r}_2}^*+b_{{\bf
        r}_2} b_{{\bf r}_1}^*) \chi_\perp(0,\bm r_4-\bm r_6)
        \chi_\perp(\Delta ,\bm r_1-\bm r_5)-\nonumber\\
        &c_{{\bf r}_4} c_{{\bf r}_6} (a_{{\bf r}_1} a_{{\bf r}_2}^*+b_{{\bf
        r}_5} b_{{\bf r}_1}^*) \chi_\perp(0,\bm r_3-\bm r_6)
        \chi_\perp(\Delta ,\bm r_2-\bm r_5)+c_{{\bf r}_1} c_{{\bf r}_6} (b_{{\bf r}_5} a_{{\bf r}_2}^*+b_{{\bf
        r}_2} a_{{\bf r}_3}^*) \chi_\perp(0,\bm r_4-\bm r_6)
        \chi_\perp(\Delta ,\bm r_3-\bm r_5)+\nonumber\\
        &c_{{\bf r}_4} c_{{\bf r}_5} (b_{{\bf r}_3} a_{{\bf r}_2}^*+b_{{\bf
        r}_6} a_{{\bf r}_3}^*) \chi_\perp(0,\bm r_1-\bm r_5)
        \chi_\perp(\Delta ,\bm r_2-\bm r_6)+c_{{\bf r}_3} c_{{\bf r}_4} (a_{{\bf r}_5} a_{{\bf r}_2}^*+b_{{\bf
        r}_6} b_{{\bf r}_1}^*) \chi_\perp(\Delta ,\bm r_1-\bm r_5)
        \chi_\perp(\Delta ,\bm r_2-\bm r_6)-\nonumber\\
        &c_{{\bf r}_1} c_{{\bf r}_5} (a_{{\bf r}_4} a_{{\bf r}_3}^*+b_{{\bf
        r}_6} b_{{\bf r}_4}^*) \chi_\perp(0,\bm r_2-\bm r_5)
        \chi_\perp(\Delta ,\bm r_3-\bm r_6)-c_{{\bf r}_1} c_{{\bf r}_4} (b_{{\bf r}_6} a_{{\bf r}_2}^*+b_{{\bf
        r}_5} a_{{\bf r}_3}^*) \chi_\perp(\Delta ,\bm r_2-\bm r_5)
        \chi_\perp(\Delta ,\bm r_3-\bm r_6)-\nonumber\\
        &c_{{\bf r}_2} c_{{\bf r}_5} (a_{{\bf r}_6} a_{{\bf r}_3}^*+b_{{\bf
        r}_3} b_{{\bf r}_4}^*) \chi_\perp(0,\bm r_1-\bm r_5)
        \chi_\perp(\Delta ,\bm r_4-\bm r_6)+c_{{\bf r}_2} c_{{\bf r}_5} (a_{{\bf r}_6} b_{{\bf r}_1}^*+a_{{\bf
        r}_1} b_{{\bf r}_4}^*) \chi_\perp(0,\bm r_3-\bm r_5)
        \chi_\perp(\Delta ,\bm r_4-\bm r_6)-\nonumber\\
        &c_{{\bf r}_2} c_{{\bf r}_3} (a_{{\bf r}_6} b_{{\bf r}_1}^*+a_{{\bf
        r}_5} b_{{\bf r}_4}^*) \chi_\perp(\Delta ,\bm r_1-\bm r_5)
        \chi_\perp(\Delta ,\bm r_4-\bm r_6)+c_{{\bf r}_1} c_{{\bf r}_2} (a_{{\bf r}_6} a_{{\bf r}_3}^*+b_{{\bf
        r}_5} b_{{\bf r}_4}^*) \chi_\perp(\Delta ,\bm r_3-\bm r_5)
        \chi_\perp(\Delta ,\bm r_4-\bm r_6)+\nonumber\\
        &c_{{\bf r}_5} c_{{\bf r}_6} (a_{{\bf r}_4} a_{{\bf r}_3}^*+b_{{\bf
        r}_3} b_{{\bf r}_4}^*) \chi_\perp(0,\bm r_1-\bm r_5)
        \chi_\perp(0,\bm r_2-\bm r_6)-c_{{\bf r}_5} c_{{\bf r}_6} (a_{{\bf r}_4} b_{{\bf
        r}_1}^*+a_{{\bf r}_1} b_{{\bf r}_4}^*) \chi_\perp(0,\bm
        r_2-\bm r_5) \chi_\perp(0,\bm r_3-\bm
        r_6)-\nonumber\\
        &c_{{\bf r}_5} c_{{\bf r}_6} (b_{{\bf r}_3} a_{{\bf r}_2}^*+b_{{\bf
        r}_2} a_{{\bf r}_3}^*) \chi_\perp(0,\bm r_1-\bm r_5)
        \chi_\perp(0,\bm r_4-\bm r_6)+ c_{{\bf r}_5} c_{{\bf r}_6} (a_{{\bf r}_1} a_{{\bf r}_2}^*+b_{{\bf
        r}_2} b_{{\bf r}_1}^*) \chi_\perp(0,\bm r_3-\bm r_5)
        \chi_\perp(0,\bm r_4-\bm r_6)).\nonumber
\label{eq:SY2_supp}
\end{align}
\end{widetext}

In order to obtain the estimates for relaxation and decoherence time, we
consider the ferromagnet to be in shape of infinite plane. Furthermore, we are
not aiming at performing an exact evaluation of the integrals in Eqs.
(\ref{eq:SX2_supp})-(\ref{eq:SY2_supp}), but rather at finding the lower bound for
the relaxation and decoherence times. To this end we note that $|C^{+-}(\omega,\bm
r-\bm r^\prime)|\le |C^{+-}(\omega,\bm r=0)|$ and arrive at the following
inequalities

\begin{align}
  S_{\tilde X_2}(0)&\le \frac{B^4}{8(\Delta_F-\Delta)^2}\int_{\Delta_F}^\infty d\nu
  C^{+-}(\nu)^2,\\
  S_{\tilde Y^-_2}(\Delta)&\le
  \frac{B^4}{8}\left(\frac{1}{\Delta_F}+\frac{1}{\Delta_F-\Delta}\right)^2\times\nonumber\\
  &\int_{\Delta_F+\Delta}^\infty
  d\nu C^{+-}(\nu)C^{+-}(\nu-\Delta),
  \label{eq:SX2SY2ineq_supp}
\end{align}
where we used notation $B=\int d\bm r B_{\bf r}$. Finally we arrive at the
estimates for the relaxation and decoherence times
\begin{align}
  T_1^{-1}\le&
  \frac{B^4S^2\Delta_F^2}{2D^3}\left(\frac{1}{\Delta_F}+\frac{1}{\Delta_F-\Delta}\right)^2
  f\left(\frac{\Delta}{\Delta_F},\beta\Delta_F\right),\\
  T_2^{-1}\le&\frac{B^4S^2\Delta_F^2}{4D^3}\left(\frac{1}{\Delta_F}+\frac{1}{\Delta_F-\Delta}\right)^2f\left(\frac{\Delta}{\Delta_F},\beta\Delta_F\right)+\nonumber\\
  &\frac{B^4S^2\Delta_F^2}{2D^3(\Delta_F-\Delta)^2}f\left(0,\beta\Delta_F\right).
  \label{eq:T1T2final1_supp}
\end{align}
with the function $f(x,y)$ defined as follows
\begin{equation}
  f(x,y)=\int_{1+x}^\infty dz
  \frac{\sqrt{z-1}}{e^{yz}-1}\frac{\sqrt{z-x-1}}{e^{y(z-x)}-1}.
  \label{eq:fxy1_supp}
\end{equation}

Assuming the same parameters as in the main text, we obtain decoherence times of
about $0.5$ hours, while the relaxation time is on the order of $1000$ hours. It
is worth noting that this result depends sensitively on the ratio $\Delta_F/T$,
thus if we assume a temperature of 4K, we obtain $T_1\ge200\mu s$ and $T_2\ge30\mu
s$.
\bibliography{ref}
\end{document}